# Compressive Wideband Spectrum Sensing for Fixed Frequency Spectrum Allocation

Yipeng Liu[†] , Qun Wan[†]

[†] Department of Electronic Engineering, University of Electronic Science and Technology of China (UESTC), Chengdu, 611731, China

{ liuyipeng, wanqun }@uestc.edu.cn

*Abstract* —Too high sampling rate is the bottleneck to wideband spectrum sensing for cognitive radio (CR). As the survey shows that the sensed signal has a sparse representation in frequency domain in the mass, compressed sensing (CS) can be used to transfer the sampling burden to the digital signal processor. An analog to information converter (AIC) can randomly sample the received signal with sub-Nyquist rate to obtained the random measurements. Considering that the static frequency spectrum allocation of primary radios means the bounds between different primary radios is known in advance, here we incorporate information of the spectrum boundaries between different primary user as a priori information to obtain a mixed l2/l1 norm denoising operator (MNDO). In the MNDO, the estimated power spectrum density (PSD) vector is divided into block sections with bounds corresponding different allocated primary radios. Different from previous standard l1-norm constraint on the whole PSD vector, a sum of the l2 norm of each section of the PSD vector is minimized to encourage the local grouping distribution while the sparse distribution in mass, while a relaxed constraint is used to improve the denoising performance. Simulation demonstrates that the proposed method outperforms standard sparse spectrum estimation in accuracy, denoising ability, etc.

## I. INTRODUCTION

Current static spectrum assignment policy leads to a paradox between the overcrowded spectrum and the pervasiveness of idle frequency bands [9]. To deal with the problem, a fast developing revolutionary technique named cognitive radio (CR) is proposed to achieve more flexible spectrum management policies and technologies for dynamic radio spectrum access [15]-[16].CR is defined as a paradigm for wireless communication in which either a network or a wireless node rapidly tunes its transmission or reception parameters to communicate efficiently avoiding interference to and from licensed or unlicensed users [24]. This alteration of parameters is based on the dynamical monitoring of the external and internal the radio environment, such as radio frequency spectrum, user behavior and network state. It is regarded as the next big thing in wireless communication.

Spectrum sensing is an essential function of CR. It autonomously detects the spectrum usage without harmful interference, and then the CR users can be allowed to utilize the unused primary frequency bands. Current spectrum sensing is performed by signal detection. it usually performed in two steps [1]: the first step called coarse spectrum sensing is to detect the power spectrum density (PSD) level of primary bands; and the second step, called feature detection or multi-dimensional sensing [25] [26], is to estimate other signal space allowing accessible for CR, such as direction of arrival (DOA) estimation, spread spectrum code identification, waveform identification, etc.

Coarse spectrum sensing requires fast and accurate power spectrum detection over a wideband and even ultra-wideband (UWB) for many CR missions [30]. One approach utilizes a bank of tunable narrowband bandpass filters to search one frequency band at a time. But it requires an enormous number of Radio Frequency (RF) components and bandpass filters, which lead to high cost and much time-consuming. Besides, the number of the bands is fixed and the filter range is always preset. Thus the filter bank way is not flexible. The other one is a wideband circuit using a single RF chain followed by high-speed DSP to flexibly search over multiple frequency bands concurrently [2]. It is flexible to dynamic power spectrum density. High sampling rate requirement and the resulting large number of data for processing are the major problems [3].

Recently a survey of a wide range of spectrum utilization across 6 GHz of spectrum in some palaces of New York City demonstrated that the maximum utilization of the allocated spectrum is only 13.1 %. That is also the reason that cognitive radio can be used. Thus it is reasonable that only a small part of the constituent signals will be simultaneously active at a given location and frequency band. The sparsity inherently exists in the wideband spectrum.

As the wideband spectrum is in sparse distribution, the Compressed sensing (CS) is a very promising technique to deal with the high sampling rate problems here [10][11][17][18]. CS is a new signal processing technique that can reconstruct the signal with a much fewer randomized samples than Nyquist sampling with high probability on condition that the signal has a sparse representation. The sampling measurement matrix should be incoherent with the dictionary. Orthogonal matched pursuit (OMP), least-absolute shrinkage and selection operator (LASSO) [19] and Dantzig Selector (DS) [20] are the major ways to recovering the sparse signal.

These standard algorithms reconstruct the whole sparse signal [4]-[7]. However, in the coarse wideband spectrum sensing, the boundaries between different kinds of primary users are fixed due to the static frequency allocation of primary radios. For example, the bands 1710~1755 MHz and 1805~1850 MHz are allocated to GSM1800. Previous compressed wideband spectrum sensing algorithms did not take advantage of the information of fixed frequency allocation boundaries. Besides, according to the practical monitoring, although the PSD vector is sparse in the mass, yet in some certain allocated frequency sections, they are not always sparse. For example, in a certain time and area, the frequency sections 1626.5～ 1646.5MHz and 1525.0～1545.0MHz allocated to international maritime satellite are not used, but the frequency sections allocated to GSM1800 are fully occupied. The wideband PSD vector is not only sparse, but rather in sparse clustered distribution with different length of clusters. This feature is extremely vivid in the situation that the monitored primary signals are all spread spectrum signals. And previous compressed spectrum sensing algorithms did not match this local feature.

Here analog to information converter (AIC) [10] [11] can be taken to obtain the random samples from analog signal in hardware as Fig. 1 shows. Another way can be mixed-signal approach [12]-[13]. In the signal recovery, we incorporate the frequency boundaries information and clustered sparsity to obtain a mixed l2/l1 norm denoising operator (MNDO). Different from the standard block-sparse signal recovery [21]-[22], the estimated PSD vector is divided into sections of different length with the known boundaries. Then a sum of l2 norm of each section of the PSD vector is minimized to encourage cluster in each section while keep sparse in the mass instead of previous l1-norm constraint on the whole PSD vector. Further exploiting the boundaries information of static frequency allocation and the clustered sparsity, the proposed mixed l2/l1 norm denoising operator MNDO based compressed wideband spectrum sensing scheme gets a performance enhancement against the traditional methods [4]-[7].

In the rest of the paper, the compressive sensing (CS) theory is reviewed a briefly in section II.; Section III gives the signal model; Section IV provides the proposed compressed wideband spectrum sensing methods; In section V, the performance enhancement of the proposed method was demonstrated by numerical experiment; Finally Section VI draws the conclusion and the possible further work.

## II. COMPRESSED SENSING REVIEW

Sparsity widely exists in wireless signals [19]. Considering a signal **x** can be expanded in an orthogonal complete dictionary, with the representation as

$$\mathbf{x}_{N\times1} = \mathbf{\Psi}_{N\times N}\mathbf{b}_{N\times1},$$ (1)

when most elements of the vector **b** are zeros, the signal **x** is sparse. And when the number of nonzero elements of **b** is S (S<<L<N), the signal is said to be *S*-sparse. CS provides an alternative to the well-known Shannon sampling theory. It is a framework performing non-adaptive measurement of the informative part of the signal directly on condition that the signal is sparse.

In CS, instead of measure the signal directly as Nyquist sampling, a random measurement matrix **Φ** is used to sample the signal. In matrix notation, the obtained random sample vector can be represented as:

$$\mathbf{y}_{M\times1} = \mathbf{\Phi}_{M\times N}\mathbf{x}_{N\times1},$$ (2)

The measurement matrix should satisfy the restricted isometry property (RIP) which is a condition on matrices **Φ** which provides a guarantee on the performance of **Φ** in CS. It can be stated as [17] [19]:

$$\left(1-\delta_s\right)\|\mathbf{y}\|_2^2 \leq \|\mathbf{\Phi}\mathbf{y}\|_2^2 \leq \left(1+\delta_s\right)\|\mathbf{y}\|_2^2,$$ (3)

for all *S*-sparse **y**. The restricted isometry constant $\delta_s \in (0,1)$ is defined as the smallest constant for which this property holds for all s-sparse vectors **y**.

There are three kinds of frequently used measurement matrices:

1) Non-Uniform Subsampling (NUS) or Random Subsampling matrices which are generated by choosing *M* separate rows uniformly at random from the unit matrix $\mathbf{I}_N$;

2) Matrices formed by sampling the i.i.d. entries $(\mathbf{\Phi})_{ij}$ from a white Gaussian distribution;

3) Matrices formed by sampling the i.i.d. entries $(\mathbf{\Phi})_{ij}$ from a symmetric Bernoulli

distribution and the elements are $\pm 1 \big/ \sqrt{N}$ with probability 1/2 each. The hardware performing random samplings of type 1 and 3 can refer to [10] [11].

When the RIP holds, a series of recovering algorithm can reconstruct the sparse signal. One is greedy algorithm, such as matched pursuit (MP), OMP; another group is convex programming, such as basis pursuit (BP), LASSO and Dantzig Selector (DS). DS has almost the same performance as LASSO. Both of the convex programming and greedy algorithm have advantages and disadvantages when applied to different problem scenarios. A very extensive literature has been developed that covers various modifications of both algorithms so to emphasize their strengths and neutralize their flaws. A short assessment of their differences would be that convex programming algorithm has a more reconstruction accuracy while greedy algorithm has less computing complex. And in contrast to BP, LASSO has additional denoising performance advantage.

## III. Signal Model

According to [9], we can suppose that receiver provides a wideband monitoring frequency range and the received signal $x(t)$, $0 \leqslant t \leqslant T$, just occupies parts of non-consecutive frequency bands with different PSD levels. The frequency range of interest is comprised of non-overlapping contiguous sections corresponding to the primary frequency allocation.

It can be represented as

$$x(t) = \sum_{i=1}^{N} c_i \psi_i(t), \qquad (4)$$

where $\Psi_i(t)$ is the Fourier basis functions and $c_i$ is the weighting factors indicating the PSD. In the weighting vector $\mathbf{c} = [c_1, c_2, \ldots, c_N]^T$, most of the elements are zeros, while the rest nonzero elements are distributed in sparse clutters as Fig. 2 shows.

In traditional Nyquist sampling, the time window for sensing is $t \in [0, NT_0]$. $N$ samples are needed to recover the PSD $r$ without aliasing, where $T_0$ is the Nyquist sampling duration. A digital receiver converts the continuous signal $x(t)$ to a discrete complex sequence $y_t$ of length $M$. For illustration convenience, we detail the algorithm in discrete setting as it did in [3]-[8]:

$$\mathbf{y}_t = \mathbf{A}\mathbf{x}_t, \tag{5}$$

where $\boldsymbol{x}_t$ represents an $N \times 1$ vector with elements $x_t[n] = x(t)$, $t = nT_0$, $n = 1, \dots, N$, and $\boldsymbol{A}$ is an $M \times N$ projection matrix. For example, when $\boldsymbol{A} = \boldsymbol{F}_N$, model (5) amounts to frequency domain sampling, where $\boldsymbol{F}_N$ is the $N$-point unitary discrete Fourier transform (DFT) matrix. Given the sample set $\boldsymbol{x}_t$ when $M < N$, compressive spectrum sensing can reconstruct the spectrum of $r(t)$ with the reduced amount of sampling data.

Processing $x(t)$ in such a broad band needs high sampling rate which is high cost. Besides, too many sampling measurements inevitably increase the computation burden for digital signal processors, while spectrum sensing requires a fast and accurate algorithm. Considering $\boldsymbol{x}_t$ has a sparse representation in frequency domain, we use an $M \times N$ random projection matrix $\boldsymbol{S}_c$ to sample signals as the type 1 random measurement in section 2, i.e. $\boldsymbol{y}_t = \boldsymbol{S}_c\boldsymbol{x}_t$, where $M < N$. The AIC can be used to sample the analog baseband signal $x(t)$. The AIC can be conceptually modeled as an ADC operating at Nyquist rate, followed by CS operation. Then $\boldsymbol{y}_t$ is obtained directly from continuous time signal $x(t)$ by AIC [11] [12]. Besides, it can be obtained by mixed-signal approach too [13]. The details about AIC can be found in references [11] [12]. Here we incorporate the AIC to the spectrum sensing architecture as Fig. 1.

## IV. THE STANDARD COMPRESSED WIDEBAND SPECTRUM SENSING

CS theory asserts that, if a signal has a sparse representation in a certain space, one can use the random measurement to obtain the samples and reconstruct it with overwhelming probability by optimization techniques, as stated in section 2. The required random samples for recovery are far fewer than Nyquist sampling requires.

To find the unoccupied spectrum for the secondary access, the signal in the monitoring band is down-converted to baseband and sampling the resulting analog signal through an Analog-to-Information converter that produces samples at a rate below the Nyquist rate.

Now we estimate the frequency response of $x(t)$ from the measurement $y_t$ based on the transformation equality $y_t = S_c F_N^{-1} r$, where $r$ is the $N \times 1$ frequency response vector of signal $x(t)$, $F_N$ is the $N \times N$ Fourier transform matrix, and $S_c$ is the $M \times N$ matrix which is obtained by randomizing the column indices and getting the first $M$ columns.

To illustration convenience, as in [3], Basis pursuit (BP) is used to recover the signal [31] [32]:

$$\hat{r} = \arg \min_r \|r\|_1, \tag{6a}$$

$$\text{s.t. } (S_c^T F_M^{-1}) r = y_t, \tag{6b}$$

This problem is a second order cone program (SOCP) and can therefore be solved efficiently using standard software packages.

BP finds the smallest $l1$ norm of coefficients among all the decompositions that the signal is decomposed into a superposition of dictionary elements. It is a decomposition principle based on a true global optimization.

As in practice the data has noise, while another algorithm called LASSO has superior denoising performance than the BP. it is a shrinkage and selection method for linear regression. It minimizes the usual sum of squared errors, with a bound on the sum of the absolute values of the coefficients. To get more accuracy, we can formulate the model by LASSO [33] as:

$$\hat{r} = \arg \min_r \|r\|_1, \tag{7a}$$

$$\text{s.t. } \left\| (S_c^T F_M^{-1}) r - y_t \right\|_2 \leq \varepsilon, \tag{7b}$$

where $\varepsilon$ bounds the amount of noise in the data. As it directly uses the sparsity to recover the wide frequency band spectrum by LASSO with reduced measurements, we named it as LASSO based compressive spectrum sensing. The computation of the LASSO solutions is a quadratic programming problem or more general convex optimization problem, and can be tackled by standard numerical analysis algorithms. The solution for it has well investigated [33]-[36]. A number of convex optimization software, such as cvx [27], SeDuMi [28] and Yalmip [29], can be used to solve the problem.

## V. THE PROPOSED COMPRESSED WIDEBAND SPECTRUM SENSING

The radio frequency (RF) electromagnetic spectrum, as land, water, and air, is a precious physical resource. Its usage is subject to usage limitations. In most of the countries and areas, it is the governments that regulate the frequency allocation to different operators for different kinds of electronic systems for technical and economic reasons. Currently for a licensed user, its operating spectrum is a certain band. Primary users employ frequency division multiplexing with fixed channelization. Part of the division information is known to the public. The bounds between different allocated spectrums of primary users are the known information. To enable dynamic spectrum access to enhance the spectrum utilization ratio, CR exploits the unused licensed spectrum. Thus for a certain monitoring wideband the boundaries for different kinds of radios is known in advance. Further utilizing the boundaries knowledge of the fixed primary frequency allocation and the local feature of the overall sparse PSD vector, the performance of signal reconstruction can be further enhanced.

Another feature of the monitoring wideband spectrum, as it is showed in Fig. 2, the active primary signals' operating frequencies in the wideband spectrum are distributed in sparse clusters, for in the same divided subband the radio has the same application function and environments. For example, in the city, the active primary users' frequency in the GSM900 bands is not sparse in the daytime. But according to the spectrum monitoring in practice, the PSD along a wideband is sparse in the mass. That is just the so called block-sparsity.

Previous common compressive wideband spectrum sensing does not assume any special structure on the unknown sparse signal. However in the practical application situation, the signal in addition to being sparse has another certain structure. Incorporating additional structure information would improve the recoverability potentially.

Block-sparse signal is the one whose nonzero coefficients are contained within several clusters. To exploit the block structure in case of ideally block-sparse signals, l2/l1 optimization was proposed. The l2/l1 optimization can be formulated as:

$$\min_{\mathbf{r}} \left( \sum_{i=1}^{K} \left\| \mathbf{r}_{(i-1)d_0 : id_0} \right\|_2 \right),$$
$$\text{s.t. } \left( \mathbf{S}_c^T \mathbf{F}_M^{-1} \right) \mathbf{r} = \mathbf{y}_t \tag{8}$$

where $K$ is the number of the divided subbands; and $d_0$ is the length of the divided blocks. Extensive performance evaluations and simulations demonstrated that as $d_0$ grows the algorithm significantly outperforms standard BP algorithm.

However, in the standard l2/l1 optimization, the estimated sparse signal is divided with the same block length, which mismatches the practical situation. Besides, the constraint in (8) does not incorporate the denoising function.

To further enhance the performance of compressed wideband spectrum sensing, the two a priori information can be incorporated in the signal recovery algorithm. Based on the known boundaries, the estimating PSD vector is divided into clusters with their edges in accordance with the boundaries. Then, in the standard LASSO, the minimizing standard $l$1-norm constraint on the whole PSD vector is replaced by minimizing the sum of the $l$2 norm of each divided group of the PSD vector to encourage the sparse in the mass while grouping locally. As it combines l1 and l2 norm to enforce the sparse clusters and subjects to a denoising constraint, we name it as the mixed l1 and l2 denoising operator (MDNO). It can be formulated as:

$$\min_{\mathbf{r}} \left( \|\mathbf{r}_1\|_2 + \|\mathbf{r}_2\|_2 + \cdots \|\mathbf{r}_K\|_2 \right)$$
$$\text{s.t. } \left\| \mathbf{y}_t - \mathbf{S}_c \mathbf{F}_N^{-1} \mathbf{r} \right\|_2 \leq \eta \quad , \tag{9}$$

where $\mathbf{r}_1$, $\mathbf{r}_2$, ... , $\mathbf{r}_K$ are $K$ sub-vectors of $\mathbf{r}$ corresponding to $d_1$, ... , $d_K$ being the lengths of divided sections. $\eta$ bounds the amount of noise in the data. It can be formulate as:

$$\mathbf{r} = \left( \underbrace{r_1 \quad \cdots \quad r_{d_1}}_{\mathbf{r}_1} \quad \underbrace{r_{d_1+1} \quad \cdots \quad xr_{d_2}}_{\mathbf{r}_2} \quad \cdots \quad \underbrace{x_{d_{K-1}+1} \quad \cdots \quad x_{d_K}}_{\mathbf{r}_K} \right)^T \quad , \tag{10}$$

Since the objective function in (9) is convex this is clearly a convex optimization problem. In principle this problem is solvable in polynomial time. To let the get the solution of (9), we can transform it to a more convenient form as:

$$\min_{\mathbf{r}, t_1, t_2, \dots, t_K} \left( \sum_{i=1}^{K} t_i \right)$$
$$\text{s. t. } \left\| \mathbf{r}_i \right\|_2^2 \leq t_i^2, \ t_i \geq 0, \ 1 \leq i \leq K \ , \tag{11}$$
$$\left\| \mathbf{y}_t - \mathbf{S}_c \mathbf{F}_N^{-1} \mathbf{r} \right\|_2 \leq \eta$$

To further simplify the solution, (11) can be reformulated to

$$\min_{\mathbf{r}, t_1, t_2, \dots, t_K} \left( \sum_{i=1}^{K} t_i \right)$$
$$\text{s. t. } \begin{bmatrix} t_i \mathbf{I} & \mathbf{r}_i^* \\ \mathbf{r}_i & t_i \end{bmatrix}, \ t_i \geq 0, \ 1 \leq i \leq K \ , \tag{12}$$
$$\left\| \mathbf{y}_t - \mathbf{S}_c \mathbf{F}_N^{-1} \mathbf{r} \right\|_2 \leq \eta$$

Obviously, (12) is a SOCP and can be solved by a host of numerical methods in polynomial time. Similar to the solution of (7), the optimal $\mathbf{r}$ of MNDO (9) can also be obtained efficiently using some convex programming software packages. Such as cvx [27] and SeDuMi [28], etc.

Comparing the standard BP algorithm (7) and the proposed MNDO (9), the object function of (7) is replaced by the mixed l2/l1 norm with different length of divided blocks. The proposed object function in of MNDO (9) encourages sparse grouping distribution, and it just fit the situation of wideband spectrum sensing for cognitive radio in practise.

But it is different from the standard mixed l2/l1 norm programming (8). In the standard mixed l2/l1 norm programming (8), the estimated signal is divided into blocks with the same block width which is an undefined structure parameter. If this parameter is mismatched, the performance of sparse signal reconstruction would deteriorate seriously. But in the wideband spectrum sensing situation, it is fortunately that the width and edges of different block subbands is in known in advance. This information is incorporated into the MNDO (9) to get the performance enhancement. Besides, instead of the constraint of standard mixed l2/l1 norm programming (8), the relaxed one of LASSO is chosen to allow the noise in practise.

With these two more piece of a priori information, i.e. boundaries information and sparse clustered distribution, the performance of proposed MNDO (9) would be better than common compressive wideband spectrum sensing way.

## VI. SIMULATION RESULTS

Numerical experiments are presented to illustrate the proposed MNDO based compressed wideband spectrum sensing for cognitive radio. Here we consider a based band signal with its frequency range from 0Hz to 500MHz as Fig. 2 shows. Four primary signals are located at 30MHz – 70MHz, 120MHz – 180MHz, 300MHz – 340MHz, 420MHz – 460MHz. And their corresponding PSD levels fluctuate in the range of 0.0277 - 0.1126, 0.0157 - 0.0988, 0.0588 - 0.1294, and 0.0381 - 0.1201, as Fig. 2 shows. And the primary signals with random phase are contaminated by a zero-mean additive Gaussian white noise (AGWN) which makes the signal noise ratio (SNR) be 13dB.

Here we take the contaminated signal as the received signal $x(t)$. As CS theory suggests, we sample $x(t)$ randomly at half the amount of Nyquist sampling's. The resulted sub-sample vector by the proposed structure as Fig. 1 demonstrated is denoted as $\boldsymbol{y}_t$. To make contrast, with the same number of samples, power spectrums with different methods are given out in Fig. 3. Fig. 3 (a) shows the result from standard LASSO based sparse spectrum recovery as (7) where $\varepsilon$ is chosen to be $0.1\|\boldsymbol{y}_t\|_2$ with 1000 tries averaged. And Fig. 3 (b) does it by the proposed MNDO as (9) where $\eta$ is chosen to be $0.5\|\boldsymbol{y}_t\|_2$; Fig. 3 shows that the MNDO gives a better reconstruction performance with the same amount of samples. To make details, it show that there are too many fake spectrum points in the subbands without active primary signals presented in Fig 3(a) for the standard LASSO. The noise levels of results from LASSO are quite high along the whole monitoring band. But for MNDO, the four occupied bands clearly show up; the noise levels in the inactive bands are quite low; and the variation of the PSD levels in the boundaries of estimated spectrum are quite abrupt and correctly in accordance with the generated sparse spectrum, which would enhance the edge detection performance much. Therefore, the proposed MNDO outperforms the standard LASSO with a much more accurate spectrum reconstruction for compressive wideband spectrum sensing as showed in Fig. 3.

Apart from the edge detection, energy detection is the most popular spectrum sensing approach for cognitive radio. To test compressive wideband spectrum sensing by energy detection, 1000 Monte Carlo simulations is done with same parameters above to give the results of average energy in each section of the divided spectrum vector with standard LASSO and the proposed MNDO. Here the simulated monitoring band is divided into 9 sections as Fig 2. The total energy with each recovering method is normalized. Table 1 presents the average energy in each subband with the corresponding recovering methods. For the MNDO, it is obvious that the estimated noise energy of inactive bands is much smaller that the standard LASSO. To quantify the MNDO's performance gain against standard LASSO, after normalizing the total energy with each method, we define the energy betterment ratio (EBR) for the $k$-th subband as:

$$EBR(k) = \begin{cases} \dfrac{\left\| \mathbf{r}_k^N \right\|_2 - \left\| \mathbf{r}_k^S \right\|_2}{\left\| \mathbf{r}_k^S \right\|_2}, & \text{for active subbands} \\[3ex] \dfrac{\left\| \mathbf{r}_k^S \right\|_2 - \left\| \mathbf{r}_k^N \right\|_2}{\left\| \mathbf{r}_k^S \right\|_2}, & \text{for inactive subbands} \end{cases}, \tag{13}$$

where $\mathbf{r}_k^N$ represents PSD vector in the $k$-th subband range obtained by the newly proposed MNDO, and $\mathbf{r}_k^S$ represents PSD vector in the $k$-th subband range obtained by the standard LASSO.

This performance function can quantify how much energy increased to enhance the probability of correct energy detection of the active primary bands and how denoising performance is enhanced. The EBR in Table 1 clearly tells the improvement of the proposed method. Although there are two subbands with declined energy, the other two active subbands have a large energy enhancement and the denoising performance in inactive subbands is improved much.

In 1000 Monte Carlo simulations, on our hardware condition, the average computing periods for the standard LASSO and the MNDO are 51.4219 seconds, 47.2369 seconds respectively. The reduction computing periods and enhancement of spectrum estimation accuracy qualifies the proposed MNDO as an excellent candidate for compressive wideband spectrum sensing.

## VII. Conclusion

This paper addresses the wideband spectrum sensing for cognitive radio. As it estimates spectrum in a wideband range, cost can be much lower than conventional narrow band filter bank method. Too deal with the too high sampling rate requirement and too large amount of data to process, the proposed MNDO for sparse signal recovery further utilizes two more pieces of a priori information for the compressed wideband spectrum sensing for cognitive radio. Thus the required number of samples to reconstruct the sparse signal can be reduced. Comparing to previous sparse signal recovery approaches, it is a more suitable candidate for the coarse wideband spectrum sensing.

ACKNOWLEDGMENT

This work was supported in part by the National Natural Science Foundation of China under grant 60772146, the National High Technology Research and Development Program of China (863 Program) under grant 2008AA12Z306 and in part by Science Foundation of Ministry of Education of China under grant 109139.

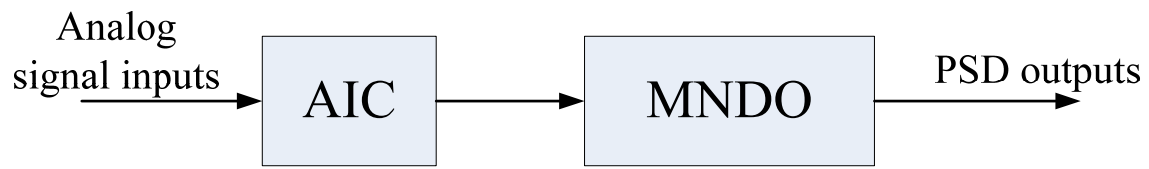

Fig. 1 The proposed compressive spectrum sensing structure

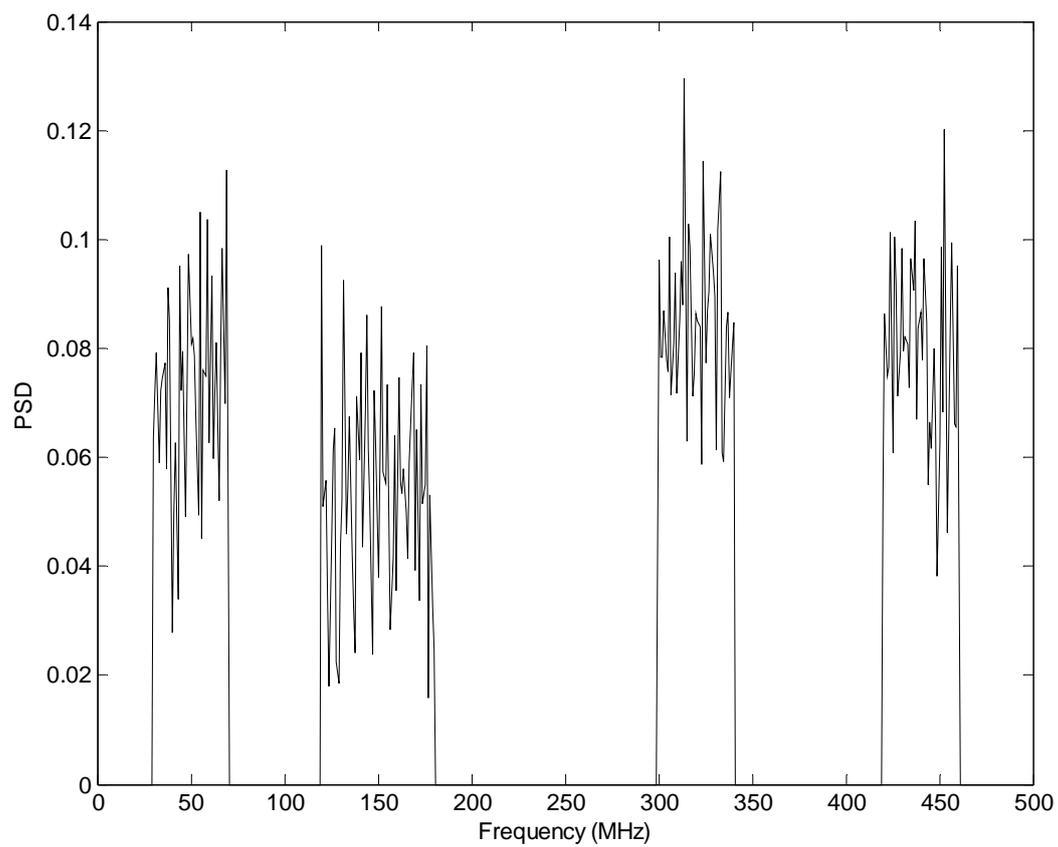

Fig 2. The spectrum of noiseless active primary signals in the monitoring band

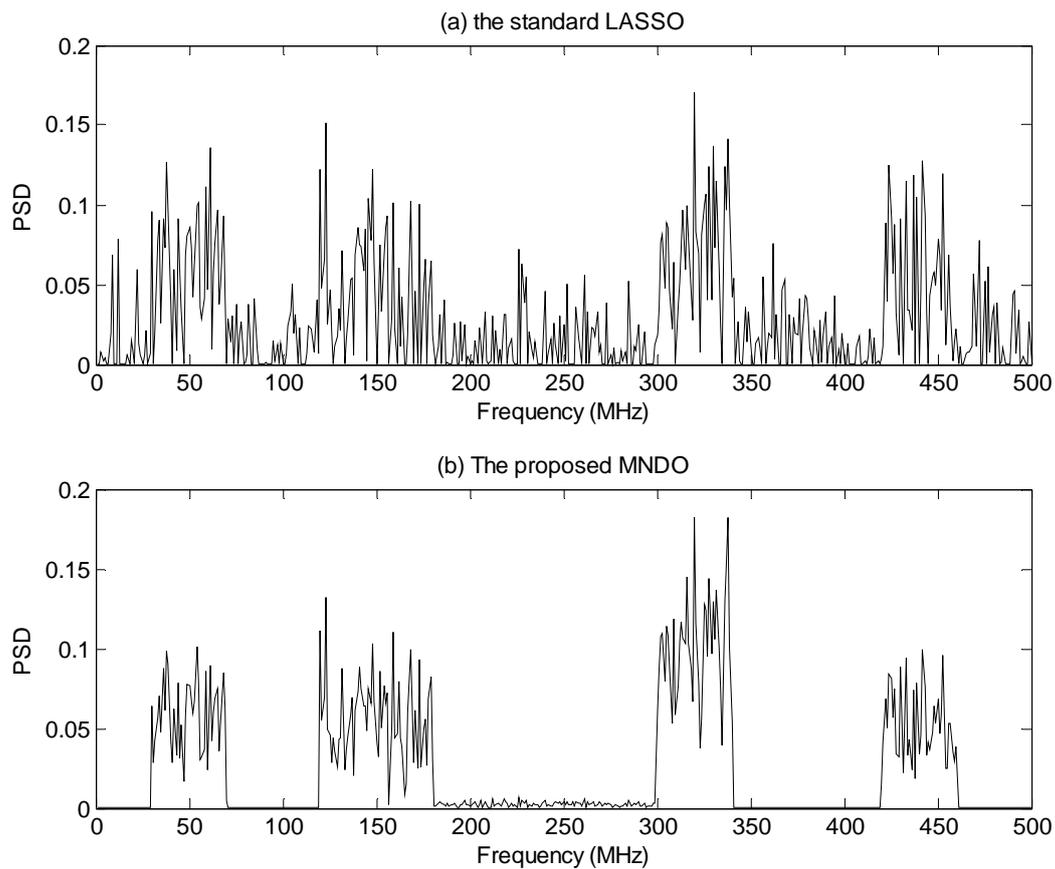

Fig. 3 The compressed wideband spectrum estimation with the standard LASSO and the proposed MNDO.

| | 1 | 2 | 3 | 4 | 5 | 6 | 7 | 8 | 9 |
|---|---|---|---|---|---|---|---|---|---|
| LASSO | 0.1269 | 0.4490 | 0.1324 | 0.4647 | 0.2165 | 0.5155 | 0.1967 | 0.4121 | 0.1666 |
| MNDO | 0.0000 | 0.3986 | 0.0000 | 0.4876 | 0.0311 | 0.6856 | 0.0000 | 0.3637 | 0.0000 |
| EBR | 100% | -11.2% | 100% | 4.93% | 85.6% | 33.0% | 100% | -11.7% | 100% |

Table 1 the total energy in each subband with two compresses wideband spectrum sensing methods and the EBR.